# Non-reciprocal heat transfer advances flexible thermoelectric devices

Jinwen Yang[1]†, Wenmei Luo[2]†, Hongbin Xu[3], Fuqing Duan[2], Yafei Ding[2], Jie Chen[4]*, Guimei Zhu[5,6]*, Baowen Li[1,2,6,7]*

This article was submitted to Science on March 17, 2026.

[1]Department of Physics, Southern University of Science and Technology, Shenzhen, 518055, People's Republic of China

[2]Department of Materials Science and Engineering, Southern University of Science and Technology, Shenzhen, 518055, People's Republic of China

[3]National Graduate College of Engineers, Southern University of Science and Technology, Shenzhen, 518055, People's Republic of China

[4]Center for Phononics and Thermal Energy Science, China–EU Joint Lab for Nanophononics, MOE Key Laboratory of Advanced Micro-structured Materials, School of Physics Science and Engineering, Tongji University, Shanghai 200092, China

[5]School of Microelectronics, Southern University of Science and Technology, Shenzhen, 518055, People's Republic of China

[6]Shenzhen Key Lab for Phononics and Intelligent Thermal Materials, Southern University of Science and Technology, Shenzhen, 518055, People's Republic of China

[7]State Key Lab for Quantum Functional Materials, Southern University of Science and Technology, Shenzhen, 518055, People's Republic of China

†These authors contributed equally to this work

*Corresponding author Email: libw@sustech.edu.cn; jie@tongji.edu.cn; zhugm@sustech.edu.cn

**Abstract:** Complex heat dissipation assemblies, inferior performance, and limited flexibility are the primary constraints impeding the wide application and commercialization of conventional flexible thermoelectric devices in wearable electronics and other high-end cooling scenarios. In this work, we report a non-conventional design for flexible thermoelectric devices which can reduce the temperature to -7.03℃ at room temperature without external heat sink, achieving a cooling temperature drop of 29.25℃. The design is based on non-reciprocal heat transfer, integrated with thermally conductive composites and screen-printing technologies. This approach takes advantage of directional heat flow, thereby eliminating the need for complex heat sink networks, which extend the applications of flexible thermoelectric devices from personal thermal management to more broader fields such as home healthcare and emergency first aid.

With the rapid development of flexible electronic devices, wearable charging and cooling technologies have attracted increasing attention. Among various wearable electronic devices, flexible thermoelectric devices (F-TED) possess unique advantages such as no mechanical components, quiet operation, and fast response *(1,2)*. They can achieve not only personalized cooling but also generate electricity by utilizing the temperature difference between the human body and the surrounding environment, demonstrating excellent potential in wearable cooling and sustainable energy applications *(3-5)*.Although many progresses have been achieved in this fast developing field, in particular, with the introduction of 3D printing and screen printing *(6-8)*, the structural designs, are still focused on two conventional types *(9-11)*: the vertical and lateral π-type structure. Both structures are vertically symmetric, leading to a bidirectional symmetry in the internal heat flow generated by the Fourier effect *(12,13)*. The maximum cooling temperature of such devices can only be maintained at approximately 15-25 ℃. For power generation, the temperature difference between the cold and hot sides of TED is usually less than 5 ℃. Consequently, to prevent the backflow of heat flow, traditional thermoelectric devices often need to attach a complex and bulky structure as heat sink to the hot end to dissipate heat *(14-16)*. The bulky heat sink, however, considerably compromises the portability, flexibility, and reliability of the devices *(17,18)*. In existing research, most researchers attempt to reduce heat flow transfer by lowering the thermal conductivity of materials. However, the Seebeck coefficient ($S$), electrical conductivity ($\sigma$), and thermal conductivity ($\kappa$) are strongly interdependent parameters, making it challenging to optimize any single one *(19)*. Reliable methods to block internal heat flow transfer while ensuring the unimpeded migration of carriers are still lacking at present.

In this work, we propose to introduce a non-symmetrical structure to F-TED. The non-symmetric structure can cause non-reciprocal heat transfer through directional thermal dissipation *(20-22)*. As it will be shown that the non-reciprocity can significantly reduce the internal energy loss in thermoelectric devices. The non-reciprocal heat transfer endows heat flow transfer with direction dependence, realizing internal heat flux blocking and significantly reducing internal energy loss caused by Fourier heat conduction. Without any external heat dissipation devices, this flexible thermoelectric device reaches a minimum cooling temperature of -7.03 ℃. This represents the maximum cooling performance reported to date for thermoelectric devices operating near room temperature *(8,23-36,38)*. Its cooling temperature difference can exceed 65 ℃, and it achieves a coefficient of performance (COP) of approximately 1.5 at a temperature difference of 29 ℃. Regarding power generation, heat flux blocking can enlarge the temperature difference, enabling the device composed of seven units to achieve potential differences of 15.6 mV and 26.1 mV during human body rest and exercise states, respectively. These performance parameters are more competitive compared with state-of-the-art TED *(8,33-37)*. Our work demonstrates the potential of non-reciprocal heat transfer in improving the performance, portability, and reliability of TED, enabling flexible TED to be used more efficiently and extensively in various cooling and energy recovery scenarios.

**Results and discussion**

Energy loss induced by internal heat flow is one of the crucial factors that degrade the performance of thermoelectric devices (16-18). In thermoelectric cooling, the cooling capacity is given by *(38-40)*:

$$Q_c = S \cdot I \cdot T_c - 0.5 I^2 \cdot R - K(T_h - T_c) \tag{1}$$

where α, R, and K are the Peltier coefficient, electric resistivity and thermal conductance, respectively. The three term $S \cdot I \cdot T_c$, $0.5 I^2 \cdot R$ , and $K(T_h - T_c)$ thus represent the Peltier effect, Joule effect and , the Fourier effect, respectively. During the operation of thermoelectric devices, carriers migrate from the cold end to the hot end, generating the Peltier effect, which creates a temperature difference between the cold and hot ends. This temperature difference, in turn, induces a reverse heat flow from the hot end to the cold end, that is Fourier effect *(41,42)*, intensifying internal energy loss and limiting further improvement in thermoelectric cooler performance *(43)*. From this equation, it is quite clear that, in order to achieve higher cooling capacity, we need to minimize the Fourier effect and the Joule effect. When an electric potential is applied to TED, carriers such as holes and electrons migrate from one end to another to generate a temperature difference. With the setup of temperature difference, heat current is generated and flows back to the cold end via thermal conduction. To quantitatively characterize the impact of Fourier heat conduction on thermoelectric cooling performance, a dimensionless parameter POF (Proportion of Fourier heat conduction) is defined as：

$$POF = \frac{K(T_\mathrm{h} - T_\mathrm{c})}{SIT_\mathrm{c}} \mid \mathrm{Qc=0} \quad (2)$$

This parameter represents the proportion of cooling capacity consumed by Fourier heat conduction relative to the Peltier effect when the cooling capacity $Q_\mathrm{c}$ is 0. The material properties and structural parameters used for the calculation are listed in Supplementary Table 1. At low input power, the proportion of Fourier heat is exceedingly high, even exceeding 90%. The variation of POF values under different temperature differences is relatively small, indicating that the heat dissipation capacity (which determines the temperature of $T_\mathrm{h}$) of TED is not the main factor affecting POF. This means that even under extreme operating conditions, enhancing heat dissipation at the hot side has a limited effect on reducing the internal energy loss caused by internal heat flow (Fig. 1a). Therefore, a new strategy is needed. Herein, we introduce a non-reciprocal heat transfer to a flexible TED (Fig. 1b). The non-reciprocal heat transfer or thermal diode was introduced 20 years ago and has been proved to be very useful for manipulating directional heat flow *(22,44, 45)*.

In this work, the asymmetric structure is designed through the alternating arrangement of bulk alloys and thin films, therefore breaking the spatial (up-down) symmetry of conventional thermoelectric devices. As shown in Fig. 1c, a thermoelectric unit consists of two p-type and two n-type bulk alloys, as well as one p-type and one n-type thermoelectric porous thin film. The porous thin film has very low thermal conductivity (typically below 1 W·m-1·K-1). Moreover, at the hot side of the TED, the heat sink is replaced by a highly conductive flexible film. With this asymmetric structure, the heat at hot end is effectively dissipated thus Th is largely reduced. And the heat flow from hot end to the cold end is also hindered because of the large thermal resistance of the porous thin film. Due to the extended heat flow path, the residual heat flow is dissipated during transmission along the thin film, causing its temperature close to the ambient temperature. This effectively blocks heat flow backward to the cold end. Consequently, the internal temperature of traditional TED shows a linear decreasing trend, while that of the TED designed based on the non-reciprocal heat transfer principle exhibits a gradient change. The high heat flow ratio results in an extremely low temperature variation across the thin film (close to the ambient temperature), thereby blocking up to over 90% of internal energy loss caused by the Fourier effect and significantly improving the thermoelectric performance (Fig. 1d, Fig. S1). For power generation, the asymmetric structure and high heat flow ratio block heat flow transfer, thereby enlarging the temperature difference to enhance power generation performance. At an input current of 2.1 A and without heat dissipation devices, the temperature of the thermoelectric device reaches -7.03 ℃, achieving a temperature drop of 29.25 ℃ relative to the ambient temperature ($T_\mathrm{amb}$=22.22℃). This value substantially surpasses the cooling performance of other flexible thermoelectric devices without heat sinks (Fig. 1e).

This extends the effective cooling temperature range of flexible thermoelectric devices from traditional daily personal thermal management to temperatures required for physical fever reduction, detumescence and analgesia for sports injuries and nerve damage, and even surgical-level ice caps (Fig. 1f). We further compared the cooling performance, heating performance, temperature difference, flexibility, power generation performance, and cost-effectiveness using different methods (Fig. 1g). The results indicate that our device is more competitive than other design methods in multiple aspects. In particular, the device's design and material fabrication methods exhibit particularly attractive potential for large-scale production. These data demonstrate that the F-TED designed based on the non-reciprocal heat transfer concept holds immense value in flexible cooling or power generation applications.

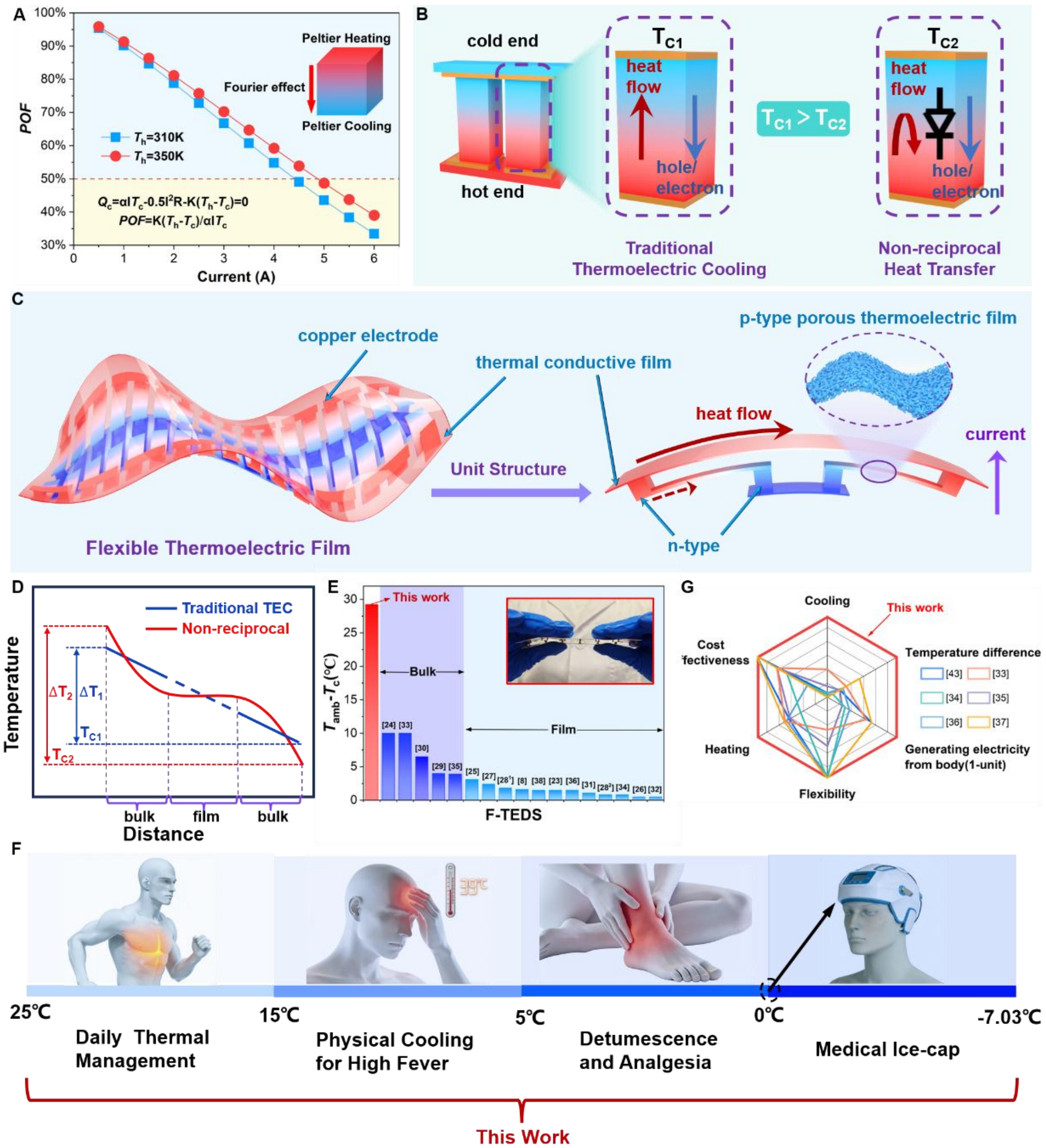

**Fig. 1 Introduction to the TED device based on non-reciprocal heat transfer. (a)** Ratio of Fourier heat to Peltier heat. **(b)** Principles and distinctions between a traditional TED and the TED based on non-reciprocal heat transfer. **(c)** Working principle of the flexible TED based on nonreciprocal heat transfer. **(d)** Comparison of temperature distribution and heat transfer between the flexible TED based on non-reciprocal heat transfer and a traditional TED. **(e)** Comparison of cooling performance without heat sinks between the TED presented in this work and those reported in references (8, 23-36, 38). **(f)** Application scope and scenarios for the flexible TED based on non-reciprocal heat transfer. **(g)** Performance comparison with references (33-37, 43) across multiple metrics: cooling, temperature difference, power generation per single element, flexibility, heating, and cost.

We evaluated the relationships between device performance and current, bulk geometry design, as well as thin film structural parameters through COMSOL simulation. The simulation methodology, including the governing equations and boundary conditions, can be found in the Methods section. The material and operational parameters used in the simulations are listed in Fig S2. The three-dimensional temperature distribution diagram of the TED simulation results is shown in Fig. S3-S5. Due to the spatially asymmetric structure of the designed TED, the Joule heat and Fourier heat conduction generated by the bulk at the hot end can be dissipated through the flexible heat dissipation interface and thermoelectric thin film. Peltier heat is generated at the interface connected to the electrodes. Compared with the low thermal conductivity of thermoelectric materials, the flexible heat dissipation interface has a higher thermal conductivity, which enables more heat to be transferred to the flexible heat dissipation thin film, endowing the heat flow transfer with a certain directionality. A small amount of heat transferred to the thermoelectric thin film is dissipated to the environment, thereby completely blocking the heat transfer.

In traditional thermoelectric coolers, increasing the cross-sectional area of the bulk improves the area-to-length ratio, which is inversely proportional to the magnitude of Joule heat and directly proportional to Fourier heat conduction. Therefore, it is necessary to optimize the cross-sectional size of the bulk to achieve the best performance *(46, 47)*. However, simulation results for the bulk structure (Fig. 2a) show that the cross-sectional area of the bulk has a very limited impact on cooling performance. This is because the equivalent thermal resistance and electrical resistance of the device mainly depend on the thermoelectric thin film, rendering changes in the bulk cross-sectional area less impactful on cooling performance. Based on the same principle, the change in the height of the bulk also has a negligible effect on cooling performance under these conditions (Fig. 2c), thus greatly reducing the structural design requirements for the thermoelectric bulk.

To demonstrate the influence of equivalent thermal resistance and electrical resistance on device performance, we increased the thermal conductivity and electrical conductivity of the connector part, and its specific material properties are shown in Table 2. Due to the increased thermal conductivity of the thin film, the equivalent thermal resistance of the entire device is significantly reduced. Although heat dissipation to the environment can mitigate some internal conduction, a portion of the heat flow still transfers to the cold side, leading to a decline in cooling performance (Fig. 2b). Therefore, when the thermal conductivity of the thin film is high, reducing the cross-sectional length of the bulk or increasing the bulk height (Fig. 2d) can direct more heat flow to the flexible heat dissipation interface. In this scenario, the bulk with high thermal resistance replaces the thin film to block heat flow, while the main role of the thin film is to extend the heat dissipation. This combination can also significantly reduce internal heat loss, improve refrigeration performance noticeably, lower the design requirements for the connector, and enhance the reliability of the device.

The structure of the thin film also has a significant impact on cooling performance. Although the thermoelectric thin film can effectively increase the equivalent thermal resistance, it also increases the equivalent electrical resistance of the TED. Therefore, when the thermoelectric thin film becomes longer and thinner, its Joule heat increases sharply. The environmental heat dissipation becomes insufficient to eliminate this excess Joule heat, resulting in a significant drop in cooling performance if all other structural parameters remain the same (Fig. 2e). In this case, however, its high electrical conductivity significantly improve the thermoelectric performance (Fig. 2f). Although the equivalent thermal resistance also decreases accordingly, the increase in its length and cross-sectional area can further extend the heat flow transfer path, thereby enhancing heat dissipation to the environment to block heat flow transfer and achieve excellent cooling performance. To show the effect of asymmetric heat flow caused by thermal non-reciprocity on performance, the heat dissipation of the connector to the environment was set to adiabatic to simulate the cooling performance of the device. Defining the thermal conductivity ratio between the heat dissipation film and the internal film as r, it is observed that device cooling performance decreases markedly at low r values. However, due to the excellent heat dissipation performance caused by spatial asymmetry, a clear marginal diminishing effect on cooling performance emerges as r increases (Fig. 2g). Therefore, the design of high thermal resistance does not require an excessively high heat flow ratio. Considering the additional environmental heat dissipation present in practical thin films, maintaining r within a range from 10 to 45 offers a more comprehensive balance between performance, flexibility, electrical insulation, and cost. Simulations were also conducted for power generation performance. Benefiting from the higher equivalent thermal resistance, under natural convection conditions, the potential difference generated by the high thermal resistance device under different thin film and bulk structures is superior to that of the low thermal resistance device (Fig. 2h and 2i). Moreover, the power generation performance of the high thermal resistance device is hardly affected by the structures of the bulk and thin film, but is more influenced by material properties.

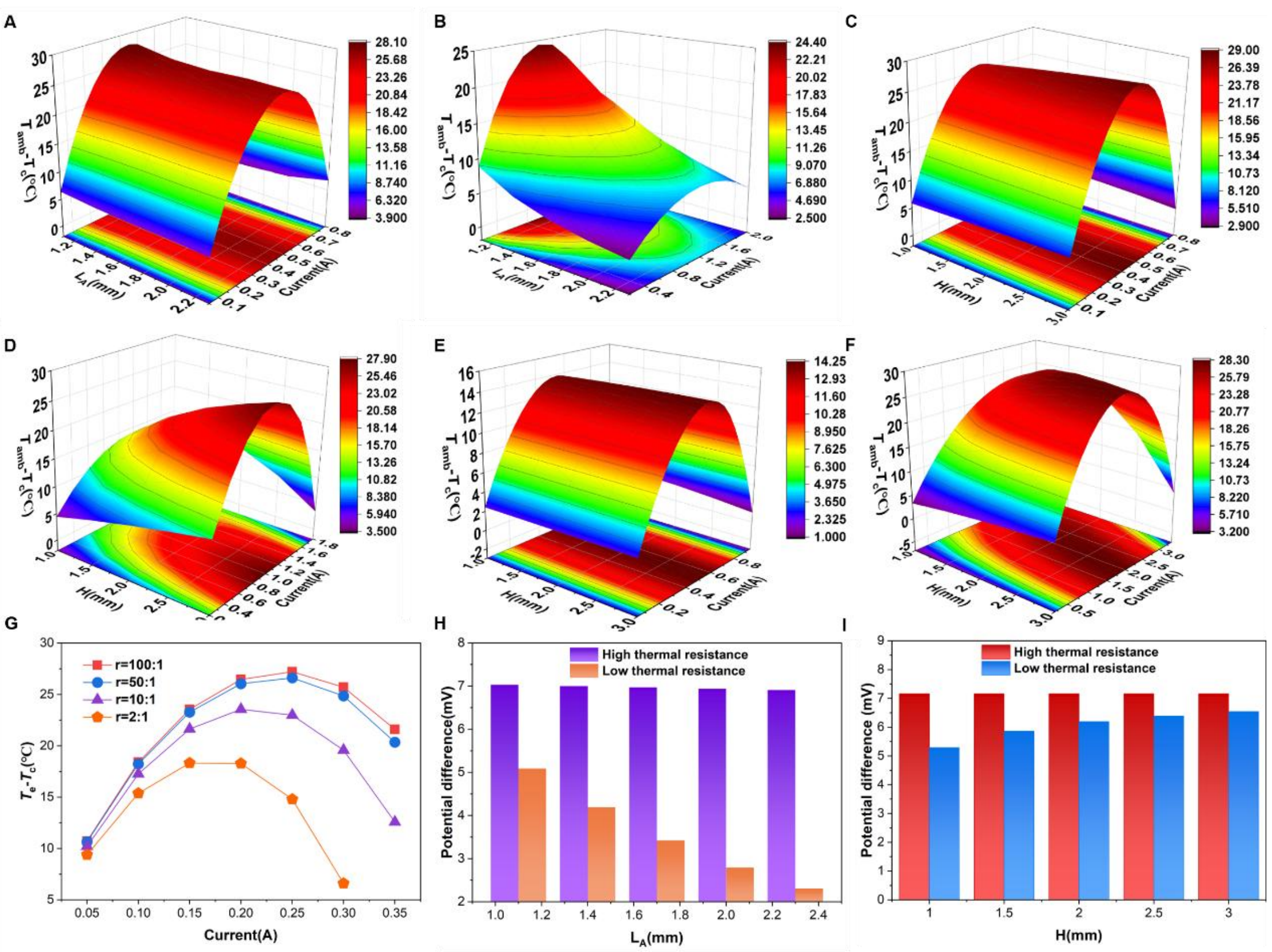

**Fig. 2 Performance simulation of the designed TED with different structures.** (a) Relationship between bulk cross-sectional width, input current, and cooling temperature when the thermoelectric thin film is 10×0.25 mm. (b) Relationship between bulk cross-sectional width, input current, and cooling temperature when the low thermal resistance thin film is 10×0.25 mm. (c) Relationship between bulk height, input current, and cooling temperature when the thermoelectric thin film is 10×0.25 mm. (d) Relationship between bulk height, input current, and cooling temperature when the low thermal resistance thin film is 10×0.25 mm. (e) Relationship between bulk cross-sectional width, input current, and cooling temperature when the thermoelectric thin film is 20×0.1 mm. (f) Relationship between bulk cross-sectional width, input current, and cooling temperature when the low thermal resistance thin film is 20×0.1 mm. (g) Relationship between the thermal conductivity ratio of the heat dissipation thin film to the thermoelectric thin film under adiabatic conditions and cooling performance. (h) Power generation performance versus bulk cross-sectional width under low and high thermal resistance thin films. (i) Power generation performance versus bulk height under low and high thermal resistance thin films.

The main functions of the thin films in the device are to reduce heat flow, connect the circuit, extend the heat flow transfer path, and ensure the flexibility of the device. Currently, the methods for thermoelectric thin films include magnetron sputtering *(48, 49)*, chemical vapor deposition *(50, 51)* (CVD), physical vapor deposition *(52, 53)*, solvothermal synthesis *(54-56)*, and screen printing *(57)* etc. Compared to other techniques, thermoelectric thin films prepared via screen printing inherently possess a porous nature (Fig. 3a) *(57)*. This porosity allows for a reduction in thermal conductivity while maintaining relatively high electrical conductivity, making them suitable for thermoelectric devices based on non-reciprocal heat transfer. Furthermore, this method offers simplicity and the potential for cost reduction. $Bi_{0.5}Sb_{1.5}Te_3$, as a p-type thermoelectric material, exhibits excellent performance and wide applications *(58-60)*, where Sb replaces Bi to form a layered structure (Fig. S6-S7). Compared with Bi, Sb has a smaller atomic radius and higher electronegativity. When Sb substitutes Bi, it causes a 1.79% lattice contraction in the a and b directions and a 1.47% contraction in the c direction of $Bi_{0.5}Sb_{1.5}Te_3$, and further alters the local electronic environment and affects the band structure. Therefore, the band gap of Sb-doped $Bi_{0.5}Sb_{1.5}Te_3$ reduces to 0.09 eV from 0.11 eV in the original $Bi_2Te_3$ (Fig. S8). Secondly, since the change in electrical conductivity is determined by the carrier concentration and mobility, the greater the carrier concentration and the higher the mobility, the greater the electrical conductivity of the system. Due to the doping of Sb, the carrier concentration is increased, which enables $Bi_{0.5}Sb_{1.5}Te_3$ to achieve a higher electrical conductivity $\sigma$. In addition, the increased material porosity introduced by screen printing adds extra phonon scattering centers, thereby reducing the thermal conductivity (κ), which is conducive to improving the overall thermoelectric figure of merit (ZT). $Ag_2Se$ is a recently discovered ductile n-type inorganic thermoelectric material with good mechanical strength and thermoelectric properties *(61-63)*. Its crystal structure is shown in Supplementary Fig. S7. For $Ag_2Se$, due to its smaller band gap and higher carrier concentration, it also results in a greater electrical conductivity $\sigma$ (Fig. S9), which is notably higher than that of $Bi_2Te_3$-based thermoelectric materials *(64)*.

$Bi_2Te_3$ has a typical layered crystal structure *(65, 66)*, thus showing superior thermoelectric performance on the (00l) plane *(67)*. However, when the thickness of the thermoelectric thin film exceeds several micrometers, the (00l) orientation is greatly reduced. Therefore, the in-plane power factor (S2σ) and ZT value of $Bi_2Te_3$-based thermoelectric thin films prepared by conventional methods are lower than those of their bulk alloys *(68)* or single-crystal nanosheets *(69)*. Due to the lack of pressure in the traditional tube furnace (Tf) annealing method, the annealed $Bi_2Te_3$ thermoelectric films exhibit a strong (015) orientation instead of the (00l) orientation *(8)*, which degrades the thermoelectric performance of the films.To address this issue, we employed a vacuum hot-pressing annealing process and compared it with the conventional tube furnace method. To investigate the phase information of the thermoelectric films prepared by the above methods, X-ray diffraction (XRD) characterization was performed (Fig. 3b). The results show that films annealed using the conventional tube furnace method exhibit a strong (015) orientation. In contrast, films prepared via vacuum hot-pressing display pronounced (00l) orientations, such as (006) and (0015). The orientation remains largely consistent across films annealed under different pressures. This superior orientation enables the films fabricated by this method to achieve ZT values between 0.96 and 1.05 at room temperature (Fig. 3c). The structure, composition, and corresponding elemental distribution of the p-type thin films were investigated using scanning electron microscopy (SEM) and energy-dispersive X-ray spectroscopy (EDS). Figure 3d shows the SEM image, while Figure 3e presents the corresponding EDS maps, including the overlapped elemental map as well as individual maps for Bi, Sb, and Te. The SEM micrograph reveals that the cross-section of the prepared thermoelectric film is dense and exhibits a distinct orientation under the applied pressure, which contributes to enhanced electrical conductivity. In contrast, the film surface possesses a certain degree of porosity (Fig. 3f), which hinders heat flow, resulting in a low thermal conductivity (κ) of 0.87 W·m-1·K-1 at 30 ℃.

The film also demonstrates good mechanical strength, achieving a bending radius of 5 mm (inset of Fig. 3f). Thereby meeting the design requirements for thermoelectric devices based on non-reciprocal heat transfer. The thermoelectric properties of the prepared $Bi_{0.5}Sb_{1.5}Te_3$ films were measured from 30 ℃ to 105 ℃. Within this temperature range, the Seebeck coefficient (S) exhibits minimal variation (only about 1.5%), while the electrical conductivity (σ) decreases by 10.7% (Fig. 3g). Consequently, the power factor (S2σ) reaches its maximum value of 29.17 μW·cm-1·K-2 at 30 ℃ (Fig. 3h). Using the same fabrication process, n-type $Bi_2Te_3$ and $Ag_2Se$ thermoelectric thin films were also prepared (Fig. S10-S13). These films exhibit similarly good ductility and low thermal conductivity (Fig. S14-15). Compared to $Bi_2Te_3$, the $Ag_2Se$ film possesses a slightly lower Seebeck coefficient (S) but a significantly higher electrical conductivity (σ), effectively reducing the equivalent resistance of the assembled device (Fig. S16-S19) . Furthermore, the Ag2Se film demonstrates even greater flexibility, achieving a bending radius of 5 mm even without a PI substrate (Fig. S20).

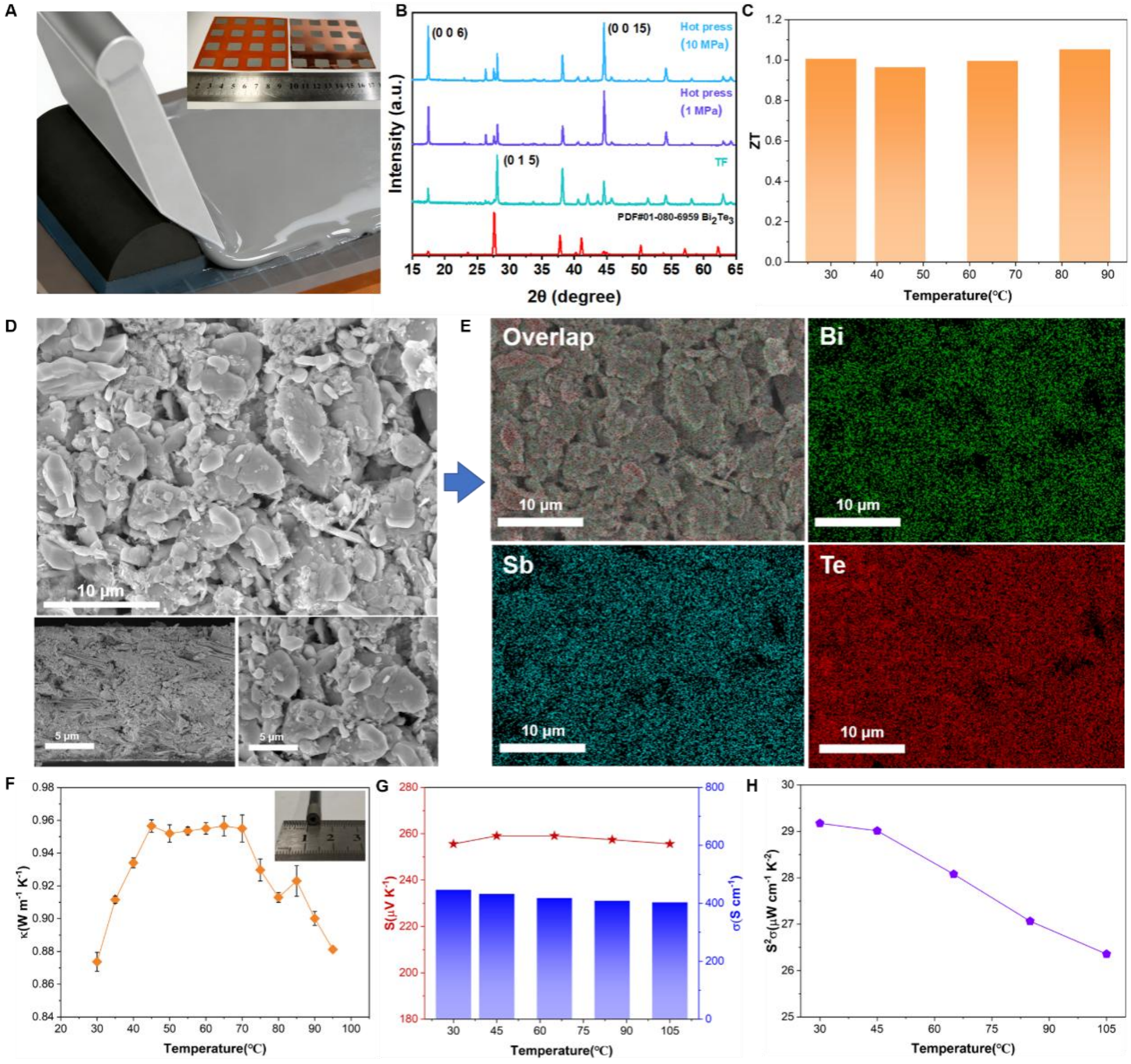

**Fig. 3 Performance and characterization of the designed thermoelectric thin films. (a)** Screen-printed thermoelectric films on PI and copper foil substrates. **(b)** XRD characterization of p-type $Bi_{0.5}Sb_{1.5}Te_3$ films subjected to different annealing methods. **(c)** ZT value of the screen-printed p-type $Bi_{0.5}Sb_{1.5}Te_3$ film. **(d)** Top-view and cross-sectional SEM images of the screen-printed p-type $Bi_{0.5}Sb_{1.5}Te_3$ film. **(e)** Corresponding EDS elemental maps. Variation of **(f)** thermal conductivity (κ), **(g)** Seebeck coefficient (S), electrical conductivity (σ), and **(h)** power factor ($S^2\sigma$) of the $Bi_{0.5}Sb_{1.5}Te_3$ film with temperature.

To ensure both flexibility and performance of the heat dissipation film, we fabricated two types of flexible heat dissipation films with excellent thermal conductivity for comparison: an anisotropic film and an isotropic film. These were made by using boron nitride and the liquid metal composite material reported in our previous work *(70, 71)* (Fig. 4a, Fig. S21), respectively, and we tested their bending performance (Fig. 4b). Non-reciprocal heat transfer devices were fabricated using these two flexible heat dissipation thin films. The results show that the performance difference between them is within 5% (Fig. S22), indicating that in-plane thermal conductivity plays a dominant role in heat dissipation. To verify the practical performance of the device, we tested the cooling performance of a single-unit device under the optimal structure, with its dimensions shown in Table 3. The infrared temperature distribution reveals that the heat transfer between the cold and hot ends is basically blocked (Fig. 4c), and this phenomenon is more obvious at low current. At an input current of 2.1 A, the temperature difference ΔT of the device reaches 49.95 ℃, with the cold end temperature and temperature drop relative to the environment being -7.03 ℃ and 29.25 ℃, respectively. This demonstrates excellent cooling performance, covering wide applications ranging from traditional personal thermal management to surgical-grade ice caps (Fig. 4d), making it competitive with existing flexible thermoelectric devices reported so far. The variation of input current alters the maximum attainable temperature difference. At an input current of 3.6 A, ΔT can reach 64.54 K (Fig. 4e). When the input current direction is reversed, the asymmetric structure leads to asymmetric heat loss, resulting in a significantly higher temperature at the hot side under the same current magnitude (Fig. 4g). This further expands the potential range for both cooling and heating applications.

In current studies, when the temperature difference between the cold and hot ends of a thermoelectric cooler reaches 25-30 K, its coefficient of performance (COP) is approximately 0.5-0.7 *(72)*. Non-reciprocal heat transfer reduces the energy loss of the device caused by internal heat flow, so the traditional COP calculation method for thermoelectric coolers cannot meet the requirements of the new device. Therefore, we calculated the COP based on energy balance. We tested the device under the optimal COP structure (its dimensions shown in Table 3), and the COP can reach approximately 1.5 at a temperature difference of 29 K (Fig. 4f), with excellent cooling performance of 5.42 ℃ (Fig. S23).

Meanwhile, we tested the power generation performance of the 7-unit device worn on the arm during standing and walking, achieving potential differences of approximately 15.6 mV and 26.1 mV, respectively (Fig. 4h). The power curves of a 2-unit device under different wind speeds were also measured (Fig. 4i) to demonstrate its power generation capability in various environments. The designed device no longer requires complex heat sinks, further improving its portability, reliability, and practicality. Overall, the outstanding performance, relatively low manufacturing cost, good flexibility, and excellent portability all together confirm the strong potential of the fabricated flexible device in practical applications.

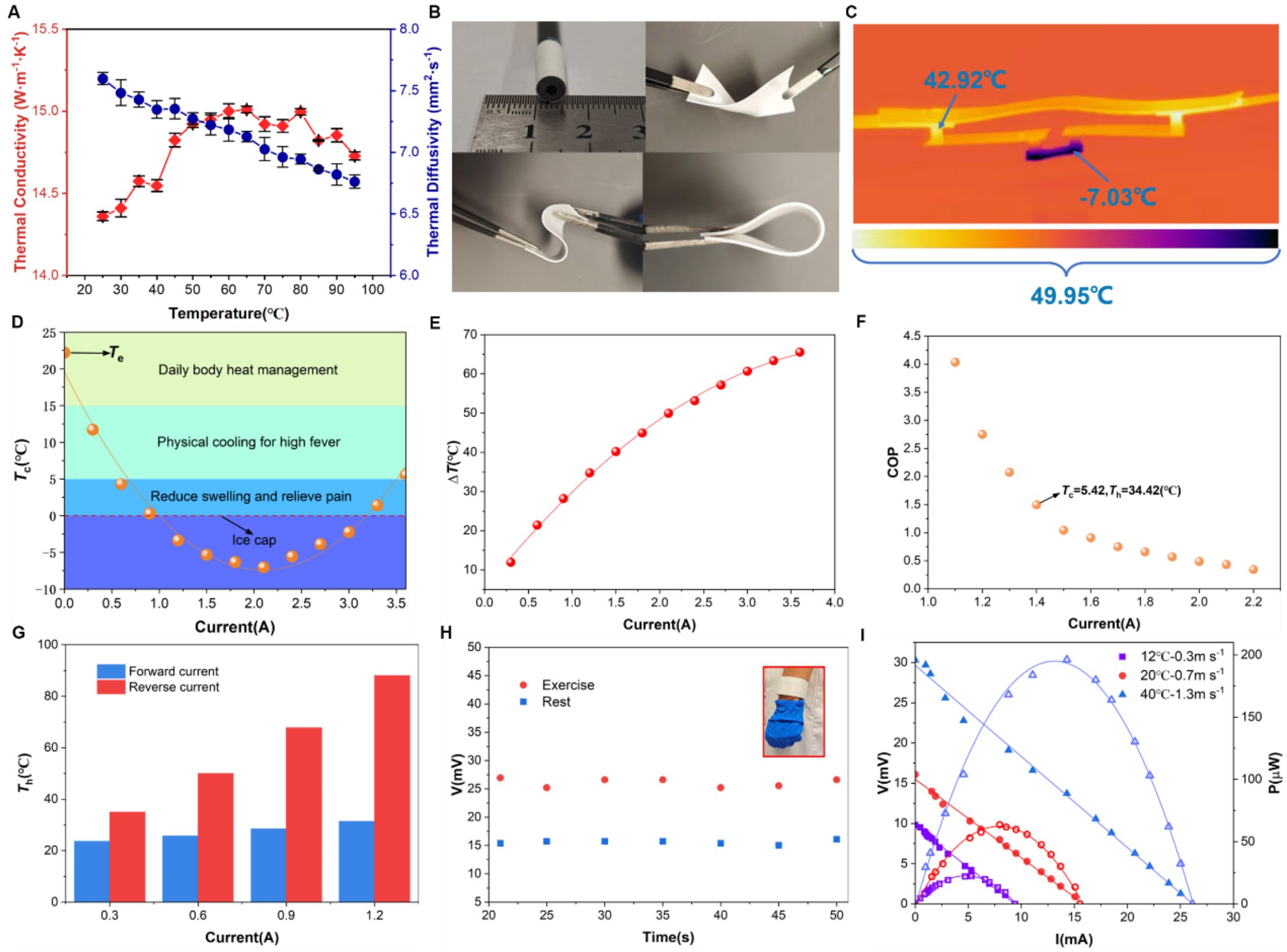

**Fig. 4 Experimental Performance of the Designed TED. (a)** Temperature dependence of thermal conductivity/diffusivity of the boron nitride thin film. **(b)** Flexibility of the boron nitride film. **(c)** Infrared temperature distribution of a single unit at an input current of 2.1 A. **(d)** Cooling temperature ($T_c$) of the TED under different input currents. **(e)** Temperature difference between the hot and cold sides ($\Delta T = T_h - T_c$) of the TED under different input currents. **(f)** COP variation of the TED with different input currents. **(g)** Comparison of heating temperature ($T_h$) after current reversal. **(h)** Output voltage (V) of the 7-unit device during walking and sitting states. **(i)** Relationships between output voltage (V), output power (P) of the 2-unit device and load current (I) under different wind speeds and temperature differences.

**Conclusions**

In summary, we have developed a high-performance flexible thermoelectric device based on non-reciprocal heat transfer, which eliminates the dependence on complex heat sinks. Thermoelectric films were fabricated via screen printing technology to further reduce internal thermal conductivity and extend the heat transfer path, while the high thermal conductivity flexible heat dissipation thin film further endows the heat flow with direction dependence. Our system exhibits excellent cooling and power generation capabilities at room temperature. It achieves a temperature reduction of 29.25 ℃ relative to the ambient temperature without any heat sink, surpassing current state-of-the-art devices of the same type (Usually around 10℃). The cold-hot temperature difference can exceed 65 ℃, and a COP of approximately 1.5 is achieved at a cold-hot temperature difference of 29 ℃. Moreover, the heat sink-free design makes the device more convenient to wear and carry. It can be widely applied globally for human thermal management, home healthcare, and surgeries in poor and under-developed regions in third-world countries. Regarding power generation, the 7-unit device can achieve a potential difference of approximately 15.6 mV under resting conditions, which further increases to about 26.1 mV during exercise. This design strategy extends the application scope of thermoelectric devices from traditional personal thermal management to a broader temperature range encompassing physical fever reduction, edema/analgesia for sports or neurological injuries, and even surgical-grade ice caps.

## Acknowledgments

### Funding:

This project has been supported by National Natural Science Foundation of China's Original Exploration Program (52250191)

Guangdong Basic and Applied Basic Research Foundation (2023ZT10X010)

Shenzhen Science and Technology In novation Committee, Grant/Award Number: JCYJ20241202125413018, SYSRD20250529114001002

Shuguang Program of Shanghai Education Development Foundation and Shanghai Municipal Education Commission (Grant No. 23SG18).

**Author contributions:**

Conceptualization: B.L, G.Z, J.C and J.Y.

Methodology: J.Y, W.L, H.X and F.D.

Investigation: J.Y, W.L and H.X.

Visualization: J.Y, W.L.

Simulations: J.Y, W.L and H.X.

Funding acquisition: B.L, G.Z and J.C.

Project administration: B.L and G.Z.

Supervision: B.L and G.Z.

Writing – original draft: J.Y and W.L.

Writing – review and editing: J.Y, W.L, H.X, F.D, Y.D, J.C, G.Z and B.L.

**Competing interests:** Authors declare that they have no competing interests.

**Data, code, and materials availability:** All data needed to evaluate the conclusions in the paper are present in the paper and the Supplementary Materials.